\documentclass[aj]{emulateapj}

\shorttitle{X-Ray Emission from the Wolf-Rayet Bubble NGC\,6888}
\shortauthors{Toal\'{a} et al.}

\usepackage{rotating}

\begin{document}

\title{X-Ray Emission from the Wolf-Rayet Bubble NGC\,6888
  \\I. \textit{Chandra} ACIS-S Observations}

\author{
J.A.\,Toal\'{a}$^{1}$, M.A.\,Guerrero$^{1}$, R.A.\,Gruendl$^{2}$, 
and Y.-H.\,Chu$^{2}$}

\affil{
$^{1}$Instituto de Astrof\'\i sica de Andaluc\'\i a,
  IAA-CSIC, Glorieta de la Astronom\'\i a s/n, 18008 Granada, Spain\\
 $^{2}$Department of Astronomy, University of Illinois, 1002 West
  Green Street, Urbana, IL 61801, US}


\begin{abstract}
  We analyze \textit{Chandra} observations of the Wolf-Rayet\,(WR)
  bubble NGC\,6888. This WR bubble presents similar spectral and
  morphological X-ray characteristics to those of S\,308, the only
  other WR bubble also showing X-ray emission. The observed spectrum
  is soft, peaking at the N {\sc vii} line emission at 0.5 keV with
  additional line emission at 0.7--0.9~keV and a weak tail of harder
  emission up to $\sim$1.5 keV. This spectrum can be described by a
  two-temperature optically thin plasma emission model ($T_{1}\sim
  1.4\times10^{6}$~K, $T_{2}\sim7.4\times10^{6}$~K). We confirm the
  results of previous X-ray observations that no noticeable
  temperature variations are detected in the nebula. The
  X-ray-emitting plasma is distributed in three apparent morphological
  components: two caps along the tips of the major axis and an extra
  contribution toward the northwest blowout not reported in previous
  analysis of the X-ray emission toward this WR nebula. Using the
  plasma model fits of the \emph{Chandra} ACIS spectra for the
  physical properties of the hot gas and the \emph{ROSAT} PSPC image
  to account for the incomplete coverage of \emph{Chandra}
  observations, we estimate a luminosity of
  $L_{\mathrm{X}}=(7.7\pm0.1)\times$10$^{33}$~erg~s$^{-1}$ for
  NGC\,6888 at a distance of 1.26~kpc. The average rms electron
  density of the X-ray-emitting gas is $\gtrsim0.4$~cm$^{-3}$ for a
  total mass $\gtrsim 1.2 M_{\odot}$.
\end{abstract}


\maketitle

\section{INTRODUCTION}

Wolf-Rayet\,(WR) bubbles are expected to be filled with hot plasma at
temperatures of $\sim10^{7}$--$10^{8}$~K , but previous X-ray
observations of hot bubbles have shown that this plasma presents lower
temperatures, of the order of $\sim$10$^{6}$~K
\citep[see][]{Chu2008}. The scarcity of X-ray detections among WR
nebulae is also intriguing; there are only two WR nebulae detected in
diffuse X-rays: S\,308 and NGC\,6888
\citep{B1988,Wrigge1994,Wrigge1998,Wrigge1999,Wrigge2002,Chu2003,Wrigge2005,Zhekov2011,Toala2012}.
These two WR bubbles share several characteristics: the X-ray-emitting
plasma is confined inside optical shells where the H$\alpha$ emission
presents a clumpy distribution inside an [O~{\sc iii}] shell
\citep{Gruendl2000}, and both bubbles are nitrogen-rich and surround
WN stars with terminal wind velocities of $\sim$1800~km~s$^{-1}$
\citep{vdH2001}. This configuration can be pictured as the WR wind
sweeping up the previously ejected Red Supergiant\,(RSG) wind material
whilst the central star photoionizing this material.

NGC\,6888 has been the subject of many studies over the years since it
was first reported by \citet{Sharpless1959} and associated to its
central WR star, WR\,136, by \citet{JH1965}. The most recent optical
study of this nebula, presented by \citet{FernandezMartin2012},
investigated the ionization, chemical composition, and kinematics in
several regions within the nebula. They concluded that NGC\,6888 is
composed by multiple shells, and its morphology can be interpreted as
a sphere with an ellipsoidal cavity inside.

The first map of the diffuse X-ray-emitting gas in NGC\,6888 was
presented by \citet{B1988} using \emph{Einstein} observations; a total
flux of $\sim$10$^{-12}$~erg~cm$^{-2}$~s$^{-1}$ was detected in the
0.2--3.0~keV band. \citet{Wrigge1994} analyzed \textit{ROSAT} PSPC
observations and found a flux of
$(1.2\pm0.5)\times$10$^{-12}$~erg~cm$^{-2}$~s$^{-1}$. \citet{Wrigge2005}
made use of the \textit{ASCA} SIS and \textit{ROSAT} PSPC observations
to fit a two-temperature model ($T_{1}\sim1.3\times10^{6}$~K,
$T_{2}\sim8.5\times10^{6}$~K) and measured a total observed flux of
$\sim$10$^{-12}$~erg~cm$^{-2}$~s$^{-1}$. The most recent X-ray
observations of NGC\,6888 are those obtained by \citet{Zhekov2011}
using the \textit{Suzaku} satellite. They concluded that the spectrum
indicates a relatively cool plasma with $T<5\times10^{6}$~K and a
small contribution from a much hotter plasma component with
temperature greater than $2\times10^{7}$~K. No appreciable temperature
variations are found between the northern and southern regions of the
nebula. The observed flux was reported to be
$2\times10^{-12}$~erg~cm$^{-2}$~s~$^{-1}$ in the 0.3--1.5~keV energy
range.

In this paper we present \textit{Chandra} observations of NGC\,6888. A
preliminary analysis of this dataset was presented by \citet{Chu2006},
where they showed the diffuse X-ray emission coming from the NE
quadrant of the nebula (ACIS-S CCD\,S3). Here we present the analysis
of the ACIS-S CCD\,S3 and CCD\,S4, covering $\sim$62\% of the
nebula. The spectral properties of the X-ray-emitting plasma are
compared to those derived previously by other authors using
observations obtained by other X-ray facilities. The data from CCD\,S4
show an additional spatial component of X-ray emission which has not
been reported in previous observations.
\\
\\

\section{\textit{Chandra} OBSERVATIONS}

The superb angular resolution and sensitivity at soft energies of
\textit{Chandra}, as compared to previous satellites that have
observed NGC\,6888, allow a more reliable study of the soft X-ray
emission from the hot plasma in this nebula. The \textit{Chandra}
observation of NGC\,6888 was performed on 2003 February 19-20
(Observation ID 3763; PI: R.A. Gruendl) using the Advanced CCD Imaging
Spectrometer (ACIS-S) for a total exposure time of 92.8~ks. The NE
quadrant of NGC\,6888 was imaged on the back-illuminated ACIS-S
CCD\,S3 while the western region was imaged by CCD\,S4. The
\textit{Chandra} Interactive Analysis of Observations\,(CIAO) software
package version 4.4 was used to analyze the data using CALDB version
3.2.2. Very short periods of high background affected the data and the
resulting useful exposure time is 88.0~ks after excising dead time
periods. The \textit{Chandra} ACIS-S observation detects diffuse
emission from NGC\,6888 in the soft energy band below 2.0~keV. No
significant emission is detected above this energy limit. The total
background-subtracted count rates of the diffuse X-ray emission for
CCD\,S3 and CCD\,S4 are 0.160 and 0.053 cnts~s$^{-1}$, respectively.

\begin{figure}[!t]
\begin{center}
\includegraphics[angle=0,width=1.0\linewidth]{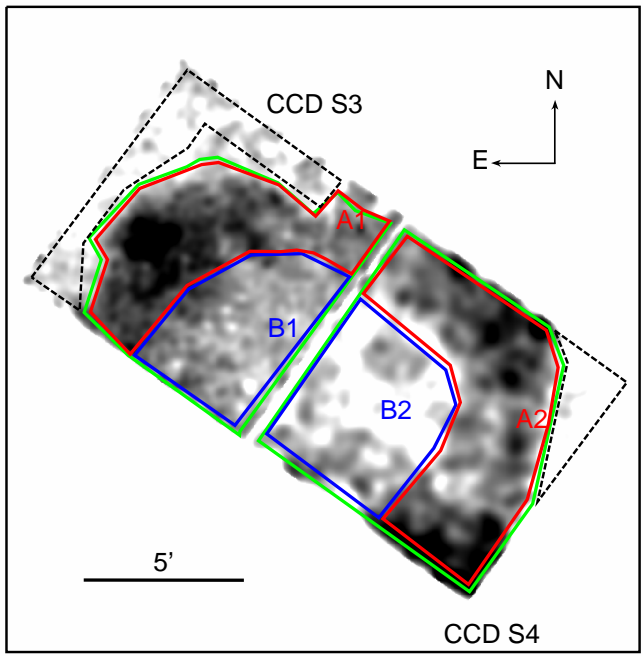}
\caption{\textit{Chandra} ACIS-S image of the diffuse X-ray emission
  of NGC\,6888 in the 0.3-2.0~keV band. Point sources have been
  excised from this image. The regions used for spectral analysis are
  indicated with polygonal apertures: 
  green, red, and blue solid lines correspond to source regions, 
  and black dashed lines to background regions.}
\end{center}
\label{fig:ngc6888_diffuse}
\end{figure}

\section{SPATIAL DISTRIBUTION OF THE DIFFUSE X-RAY EMISSION}


\begin{figure*}[!htbp]
\begin{center}
\includegraphics[angle=0,height=0.5\linewidth]{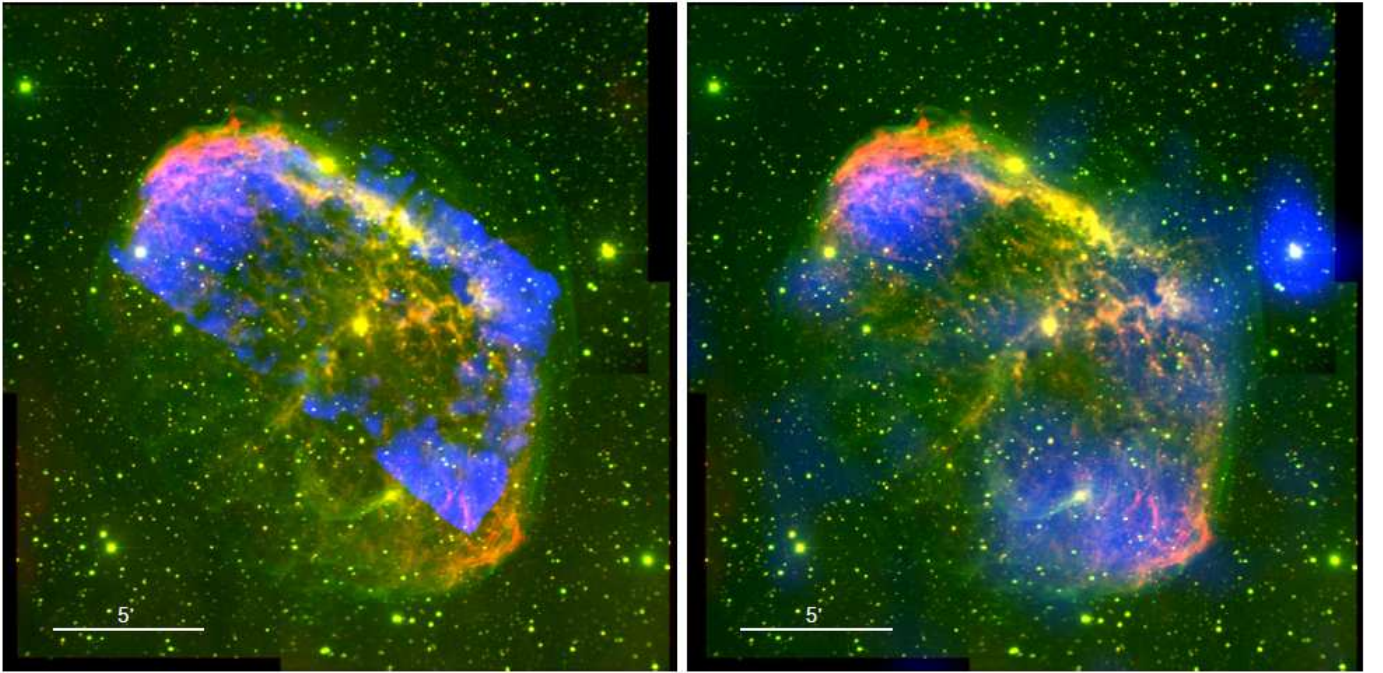}
\caption{Left: Composite color picture of the \textit{Chandra} ACIS-S
  observation of NGC\,6888 (blue) and MLO H$\alpha$ (red) and [O~{\sc
    iii}] (green) images. Right: Same as in \textit{left} image but
  with \textit{ROSAT} PSPC image (blue). North is up, East to the
  left.}
\end{center}
\label{fig:ngc6888}
\end{figure*}

In order to analyze the spatial distribution of the hot gas in
NGC\,6888, we excised all point sources from the observation using the
CIAO \textit{dmfilth} routine. The identification of the point sources
was made using the CIAO \textit{wavdetect} routine. The image of the
diffuse X-ray emission was extracted in the 0.3-2.0~keV energy band
and smoothed with the CIAO task \textit{csmooth}, with a Gaussian
kernel of 4$''$ in the brightest regions and 16$''$ and 24$''$ in the
faintest ones for the CCD\,S3 and S4, respectively. The resultant
image is shown in Figure~1.

We compare in Figure~2-left the X-ray image with H$\alpha$ and
[O\,{\sc iii}] optical images of the nebula taken with the 1~m
telescope at the Mount Laguna Observatory \citep{Gruendl2000}. This
figure shows a limb-brightened spatial distribution of the X-ray
emission confined within the optical [O\,{\sc iii}] shell. In
particular the X-ray-emitting gas in the NE region of the nebula
(ACIS-S CCD\,S3) seems to fill all the area within the nebula with a
broad emission peak superposed on the H$\alpha$ clumps, while the
emission detected in ACIS-S CCD\,S4 can be associated with the
southwest cap of the nebula, and with emission outside the H$\alpha$
shell but inside the western [O~{\sc iii}] skin.

For comparison, we also present in Figure~2-right a composite picture
of the same optical images and the X-ray emission detected by
\textit{ROSAT} PSPC \citep[][]{Wrigge1994,Wrigge2005}. This image
demonstrates that the X-ray emission from NGC\,6888 is stronger at the
caps along the major axis, but an extra contribution can be detected
at the westernmost regions of the nebula, just inside the optical
[O~{\sc iii}] shell. This additional spatial component of X-ray
emission, hinted in previous images of the nebula made with
\textit{ROSAT} HRI and \textit{ASCA}
\citep{Wrigge1994,Wrigge2002,Wrigge2005}, is reminiscent of the
Northwest blowout in S\,308 \citep{Chu2003,Toala2012}.

\section{PHYSICAL PROPERTIES OF THE HOT GAS IN NGC\,6888}

We have carried out the study of the hot gas of NGC\,6888 in several
steps.  First, we have studied the emission from the nebular gas
detected in CCD\,S3 and CCD\,S4, keeping in mind that the former has
more reliable spectral resolution and sensitivity at lower energies
than CCD\,S4.  Therefore, we have defined two regions encompassing the
diffuse X-ray emission registered in the field of view of the
\emph{Chandra} ACIS-S CCD\,S3 and S4 (green polygonal lines in
Figure~1).

For further analysis, we have defined several smaller polygonal
aperture regions, also shown in Figure~1, corresponding to different
features presented in NGC\,6888: regions labeled as A comprise the
apparent shell and caps, and regions labeled with B correspond to the
shell interior. We note that both regions are present within each CCD
detector, and thus we have extracted two spectra corresponding to each
morphological feature. For example, A1 corresponds to a region
extracted from CCD\,S3 and A2 to a region extracted from CCD\,S4. The
same applies to regions B1 and B2.

\subsection{Spectral Properties}

As discussed in \citet{Toala2012}, the extraction of spectra from
extended sources, as is the case of WR bubbles, is challenging because
the emission fills almost the entire field of view of the
instrument. 
The background contribution can be estimated from high signal-to-noise
blank fields, but as mentioned by \citet{Toala2012}, this technique
does not produce suitable results because WR bubbles are located in
regions close to the Galactic Plane where extinction and background
emission are significant \citep{Snowden1997}. To show the contribution
of the Galactic background, we plot in Figure~3 the
background-unsubtracted spectrum from CCD\,S3 and a background
spectrum extracted from the edge of the camera. X-ray emission from
the background is soft and show lines in the 0.3--1.0~keV energy band
from thermal components \citep[see also figure~5 in][for the case of
S\,308]{Toala2012}. The spectral shape of the background emission
certainly differs from that derived from ACIS blank-field
observations. Therefore, the most feasible procedure to subtract the
background contribution is the use of background spectra extracted
from areas near the camera edges, even though the instrumental
responses for sources and background regions do not completely match
each other.

\begin{figure}[t]
\begin{center}
\includegraphics[angle=0,width=1.05\linewidth]{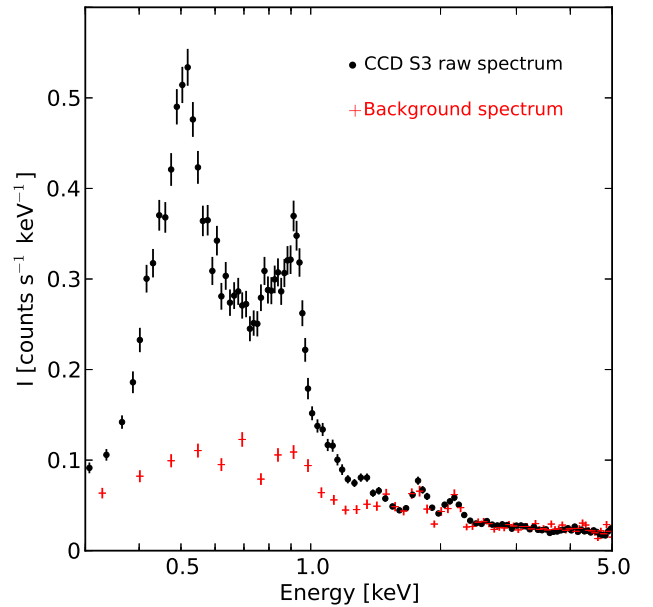}
\caption{
Comparison of the raw CCD\,S3 spectrum (black circles) and scaled 
background spectrum extracted from the edges of CCD\,S3 (red crosses). 
The emission lines around 2~keV in the background spectrum 
are instrumental lines.
}
\end{center}
\label{fig:spec_raw}
\end{figure}

The individual background-subtracted spectra of the diffuse emission
of the NE quadrant and western region of NGC\,6888 (namely CCD\,S3 and
CCD\,S4) are presented in Figure~4, as well as the individual spectra
extracted from regions A1, A2, B1, and B2. All spectra were extracted
using the CIAO task \textit{specextract}, which generates the source
and background spectra and their corresponding calibration files. The
most notable differences in the spectral shapes are attributed to the
differences in sensitivity of the ACIS-S CCD\,S3 and CCD\,S4. All
spectra are soft and show two main peaks, a narrow peak at 0.5~keV and
a broader peak around 0.7--0.9~keV. The feature around 0.5~keV can be
identified with the N~{\sc vii} ion, while the feature around
$\sim$0.7--0.9~keV can be associated with the Fe complex and Ne
lines. Above 1.0 keV, the emission declines and diminishes at energies
$\simeq$1.5 keV. We note that the spectra extracted from CCD\,S4 show
the instrumental Au M complex at 2.2~keV, which has not been properly
removed due to the reduced spatial extent of the background region.

\begin{figure*}[t]
\begin{center}
\includegraphics[angle=0,width=0.33\linewidth]{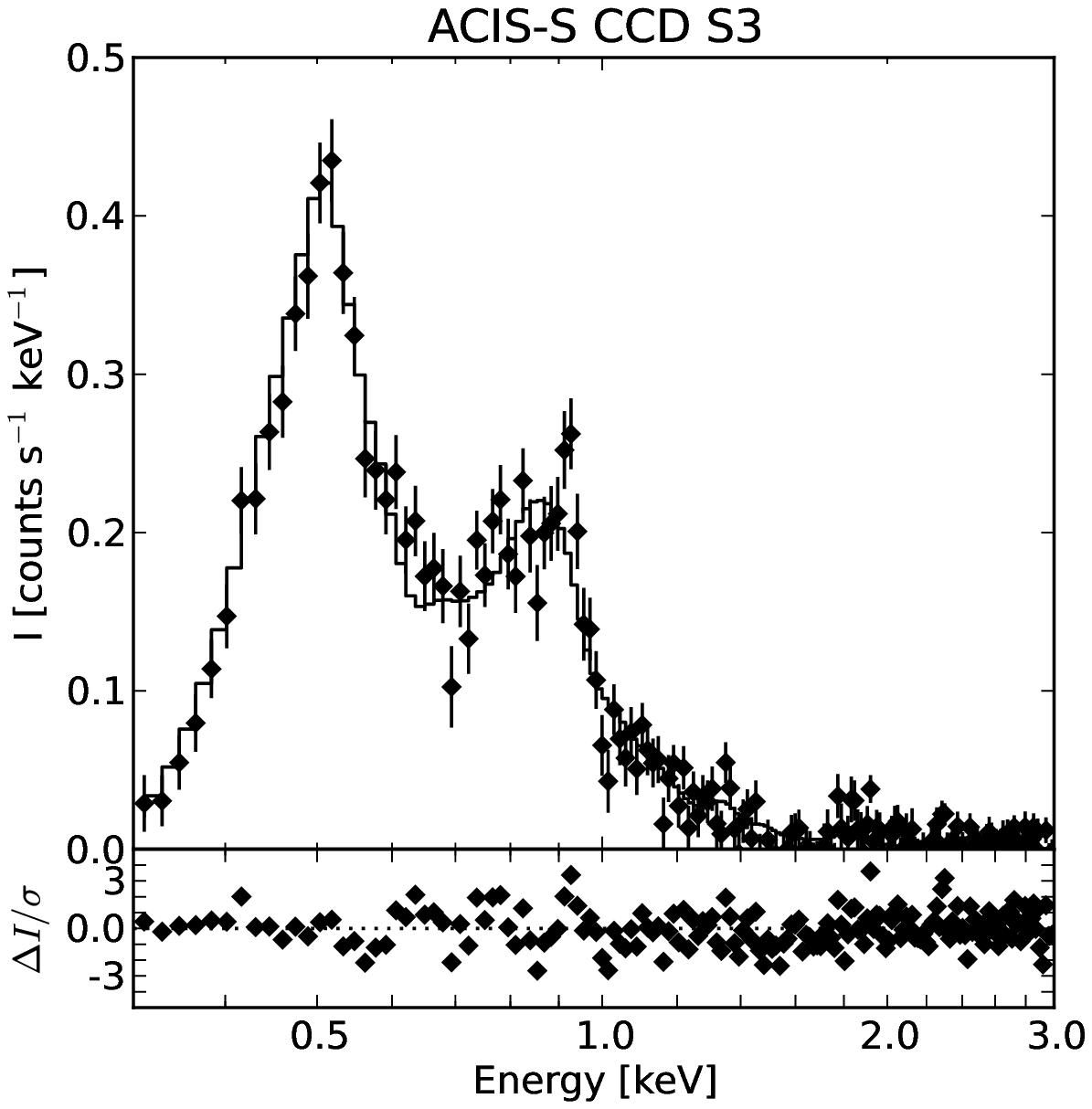}~
\includegraphics[angle=0,width=0.33\linewidth]{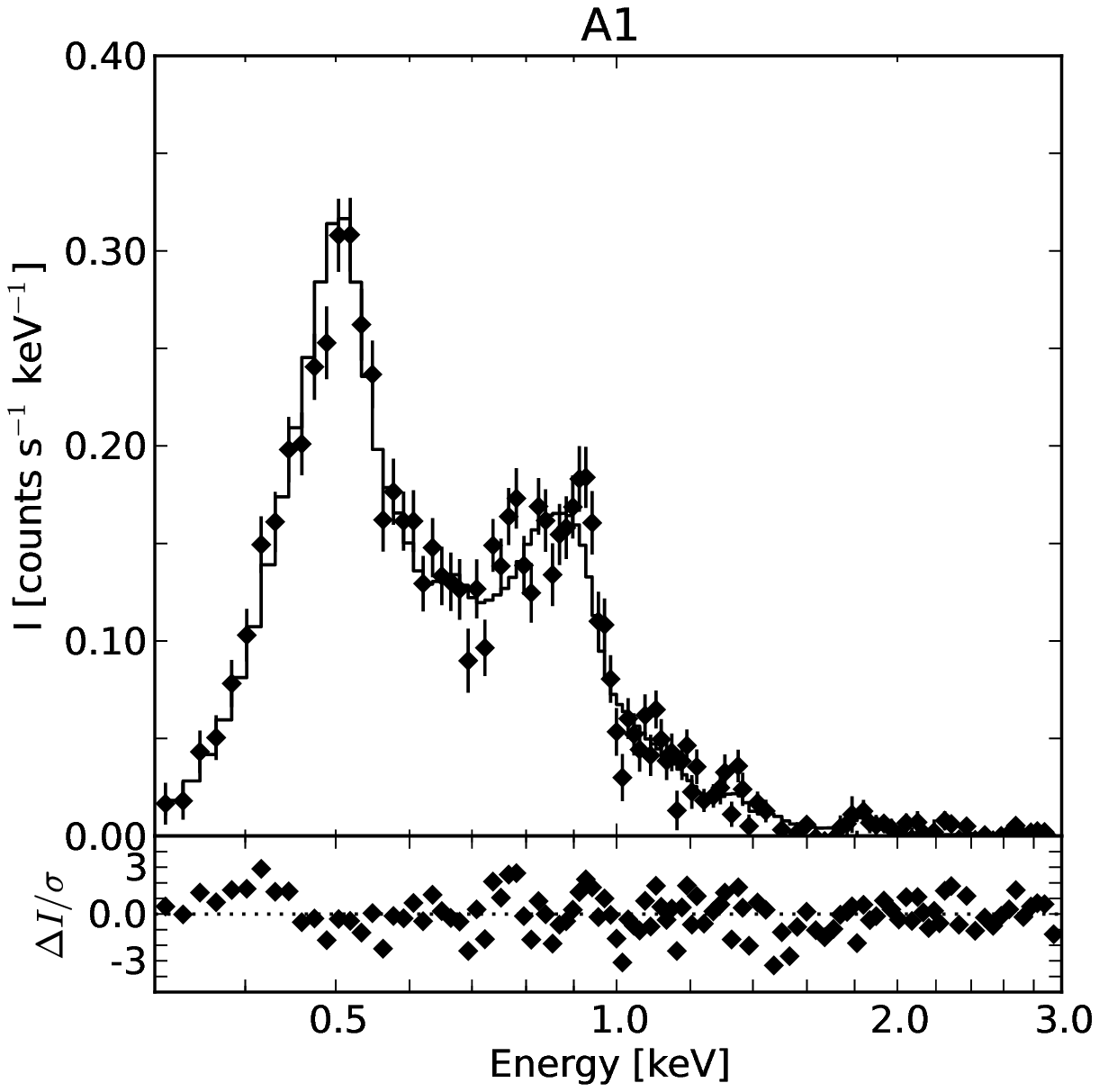}~
\includegraphics[angle=0,width=0.33\linewidth]{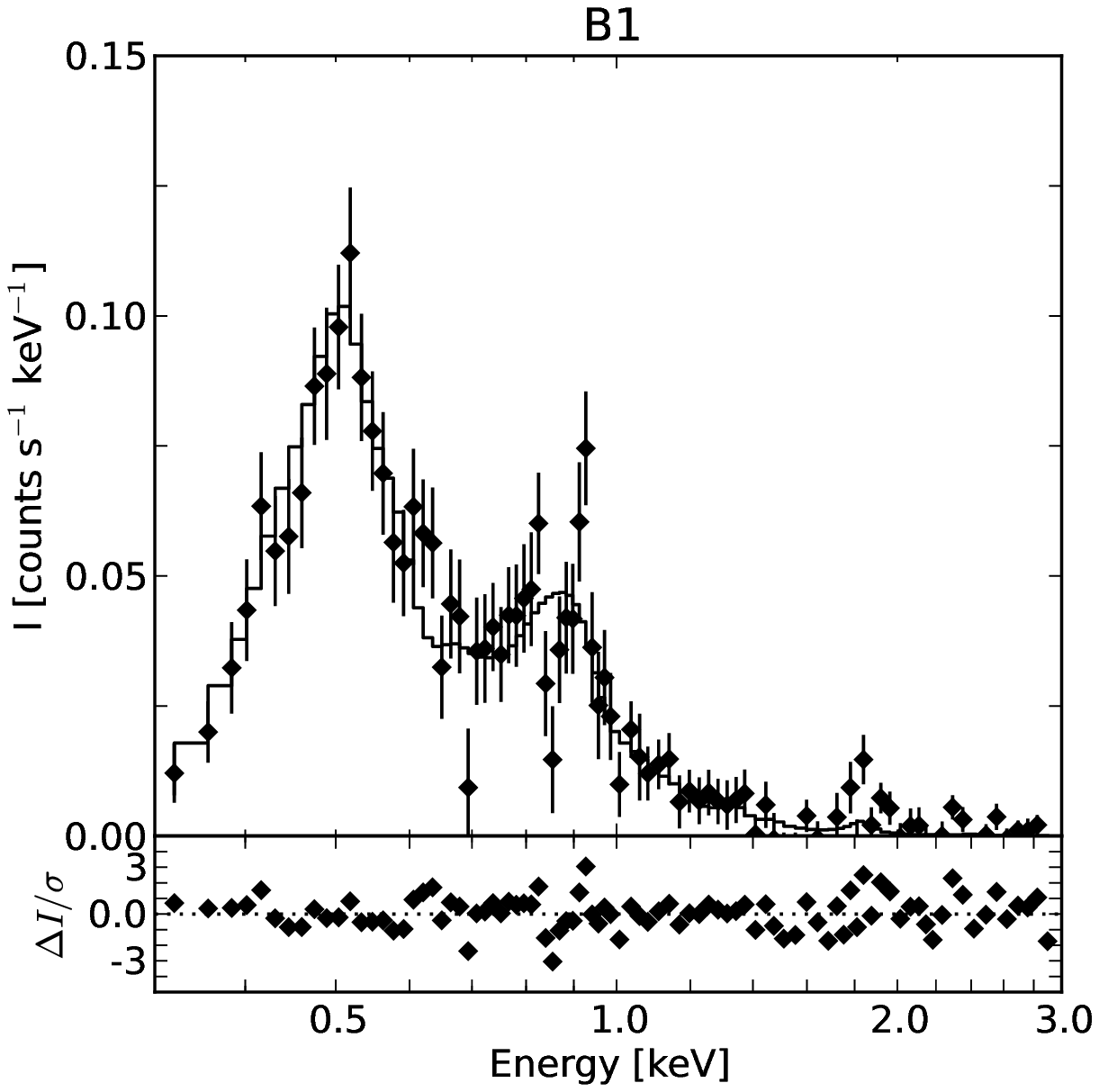}\\
\includegraphics[angle=0,width=0.33\linewidth]{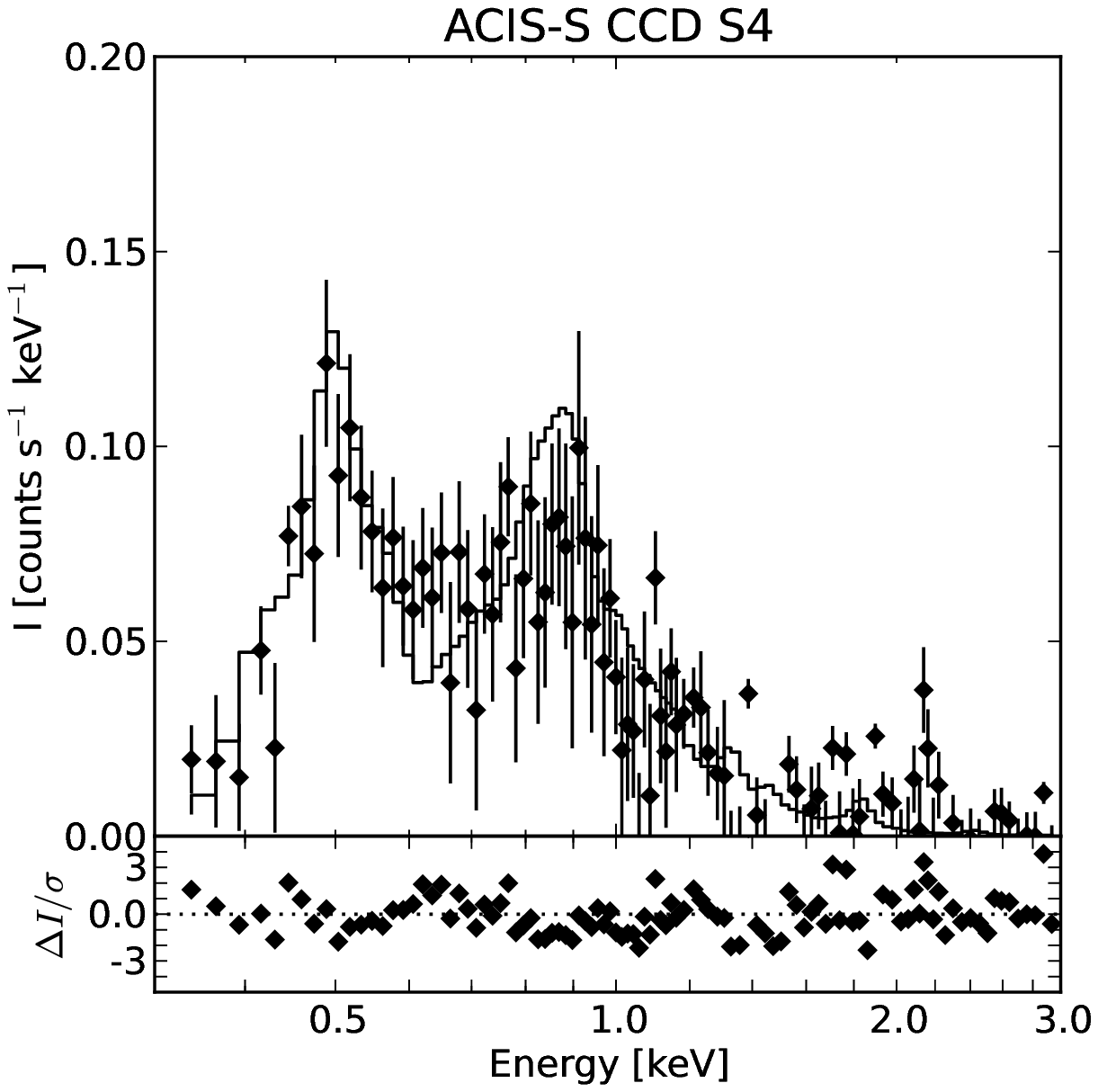}~
\includegraphics[angle=0,width=0.33\linewidth]{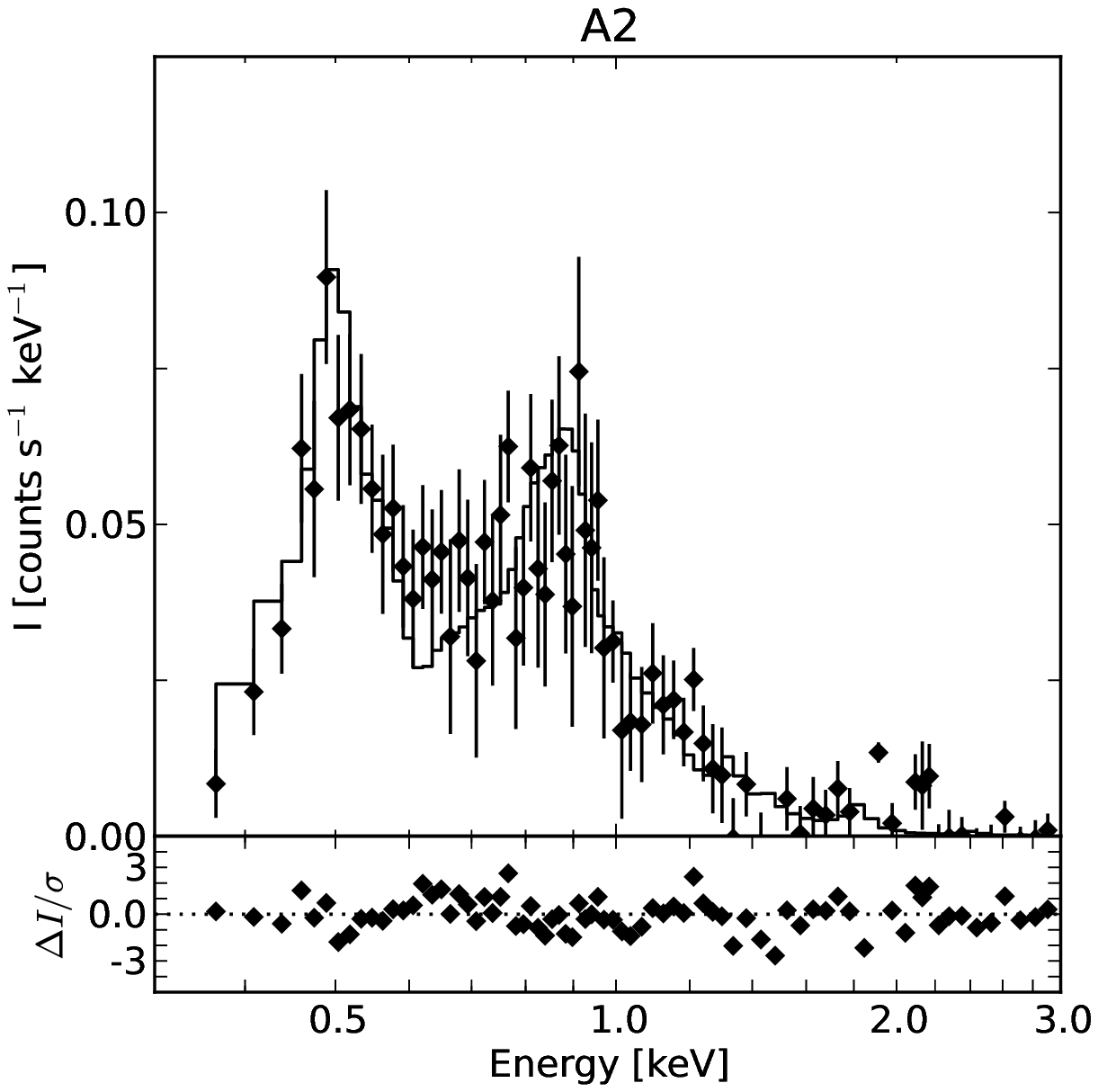}~
\includegraphics[angle=0,width=0.33\linewidth]{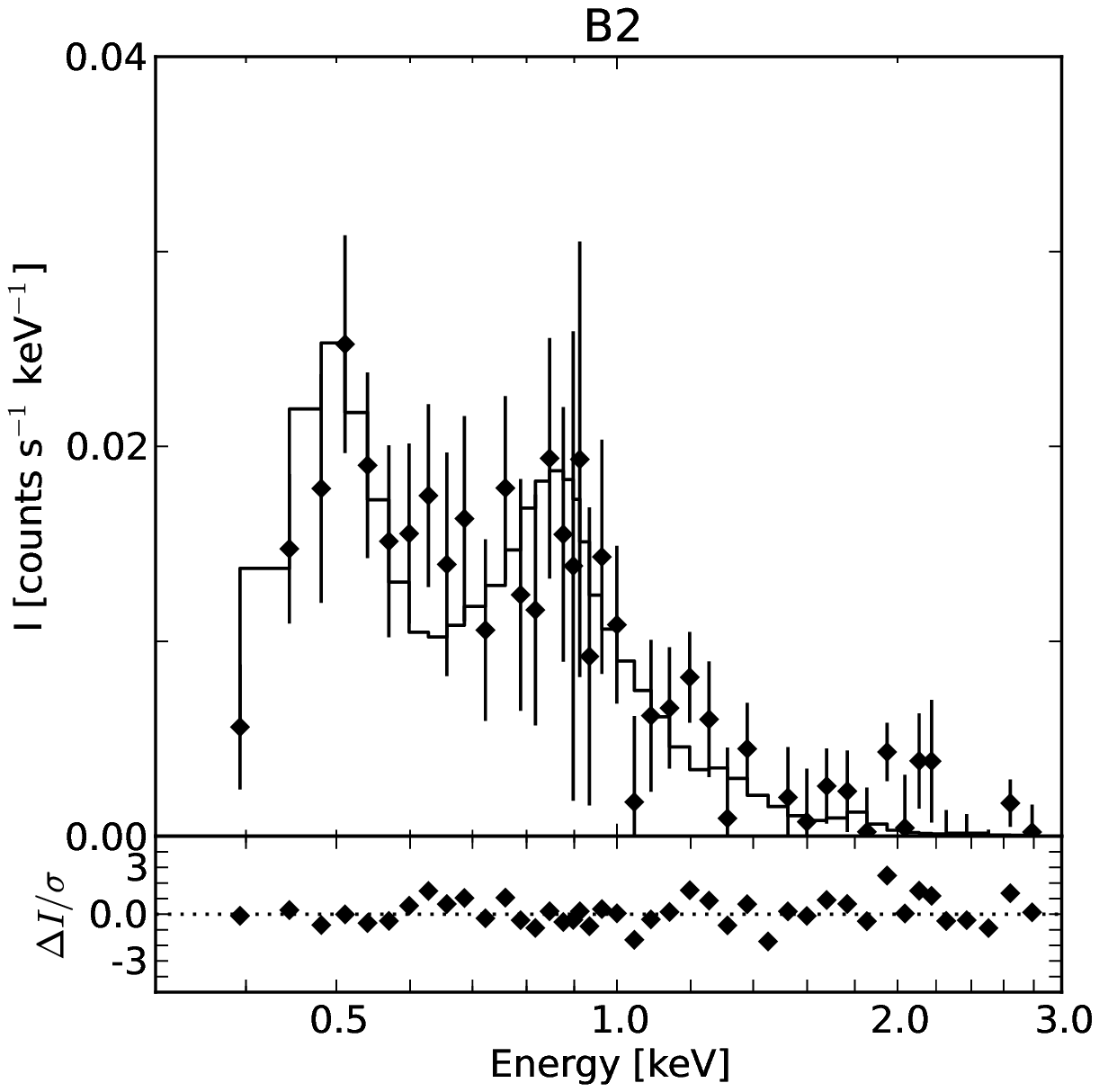}
\caption{Background-subtracted \textit{Chandra} ACIS-S spectra of the
  NE (top panels) and western (bottom panels) regions of NGC\,6888
  over-plotted with their best-fit two-temperature \textit{apec} model
  (solid lines) in the energy range of $0.3-3$~keV.  
}
\end{center}
\label{fig:spec_all}
\end{figure*}

In accordance with the spectral properties and previous spectral fits,
all X-ray spectra from NGC\,6888 have been fit with XSPEC v12.7.0
\citep{Arnaud1996} using an absorbed two-temperature \textit{apec}
optically thin plasma emission model with an absorption model using
\citet{Balu1992} cross-sections. A low temperature component is used
to model the bulk of the X-ray emission, while a high temperature
component is added to model the extra emission at and above
$\sim0.7$~keV.  As in previous studies of the X-ray emission from WR
bubbles \citep[see][]{Chu2003,Zhekov2011,Toala2012}, we have initially
adopted nebular abundances for the X-ray-emitting plasma.  In
particular, we have used abundance values for N, O, and Ne of 3.2,
0.41, and 0.85 times the solar values \citep{Anders1989} as averaged
from regions X1 and X2 described in \citet[][see their
Table~4]{FernandezMartin2012}, and 0.39 times the solar value for S
\citep{Moore2000}.  Models with variable C, Mg, Fe, and Ne abundances
were also tested.  We found that the fitted abundances of Mg and Fe
converged to solar values, whereas those of Ne tended to 0.85 times
the solar value, i.e., the value determined from optical
spectrophotometry \citep{FernandezMartin2012}.  Consequently, we
decided to fix the abundances of Mg, Fe, and Ne to these values.  As
for the carbon abundance, the fits could not converge to specific
values because the C~{\sc vi} line at 0.37~keV or C~{\sc v} triplet at
0.3~keV are in the low energy range, where absorption is high and the
instrument sensitivity is low. Therefore, we fixed the value of the
carbon abundance to its solar value. Finally, all spectra were modeled
with varying nitrogen abundance ($\mathrm{X_{N}}$) as the prominent
N~{\sc vii} line at 0.5 keV seems to suggest a possible nitrogen
enrichment of the X-ray-emitting plasma.

The simulated two-temperature \textit{apec} model spectra obtained
were absorbed by the interstellar hydrogen column of
3.13$\times10^{21}$~cm$^{-2}$ implied by optical measurements
\citep{Hamann1994}.  This is the same value used by previous authors
\citep{Wrigge1994,Wrigge2005} which was found to be uniform throughout
NGC\,6888 by \citet{Wendker1975}.  Models with variable columnar
density $N_{\mathrm{H}}$ were also attempted. The absorption column
density showed a correlation with the temperature of the main plasma
component ($T_{1}$), with values 2--4$\times10^{21}$~cm$^{-2}$ which
is adopted here and is consistent with the value used by
\citet{Zhekov2011}.

The resultant model spectra were compared with the observed spectra in
the 0.3 - 3~keV energy range and the $\chi^{2}$ statistics was used to
determine the best-fit models. A minimum of 60 counts per bin was
required for the spectral fit. The plasma temperatures ($kT_{1}$,
$kT_{2}$) with 1$\sigma$ uncertainties, normalization
factors\footnote{$A = 1 \times10^{-14}\int n_{\mathrm{e}}
  n_{\mathrm{H}} dV/ 4 \pi d^{2}$, where $d$ is the distance,
  $n_{\mathrm{e}}$ is the electron density , and $V$ the volume in cgs
  units.} ($A_{1}$, $A_{2}$), and nitrogen abundance
($X_{\mathrm{N}}$) of the best-fit models are listed in Table 1.
Fluxes and luminosities listed in this table have been computed for
the energy range 0.3-2.0 keV.  The best-fit models are over-plotted on
the background-subtracted spectra, together with the residuals of the
fits as solid lines in Figure~4.

\subsubsection{Properties of the NE X-ray Emission}

The parameters of the best-fit model of the NE quadrant of NGC\,6888
are listed in Table\,1 as CCD\,S3. The model presents a
low-temperature component of $1.6\times10^{6}$~K and a second
component of $7.8\times10^{6}$~K with an observed flux ratio,
$f_{1}/f_{2}\sim3$, corresponding to an intrinsic flux ratio
$F_{1}/F_{2}\sim14$. The total observed flux in the 0.3--2~keV energy
band is $(6.4^{+0.1}_{-0.2}) \times10^{-13}$~erg~cm$^{-2}$~s$^{-1}$
while the total unabsorbed flux is
$(8.0^{+0.4}_{-0.2})\times10^{-12}$~erg~cm$^{-2}$~s$^{-1}$. The
nitrogen abundance of the best-fit model is $\approx$5.3 times that
of the solar value.

In the case of the resultant spectra from regions A1 and B1, the
temperatures are consistent with those obtained from the whole region,
$T_{1}=1.6\times10^{6}$~K and $T_{2}=7.9\times10^{6}$~K for region A1,
and $T_{1}=1.4\times10^{6}$~K and $T_{2}=7.7\times10^{6}$~K for region
B1, respectively. Their nitrogen abundances are 5.3 and 3.7 times the
solar value for A1 and B1, respectively.

\subsubsection{Properties of the Western X-ray Emission}

The background-subtracted X-ray spectra from the western regions of
NGC\,6888 are shown in the lower panels of Figure~4, which correspond
to the total diffuse emission detected by the CCD\,S4 (left), the
emission from the rim registered by region A2 (middle), and the
emission from inside the WR nebula registered by region B2 (right).
The parameters of the best-fit models over-plotted to these spectra
are presented in Table~1.  We remark that the fits of these spectra
failed to constrain accurately the nitrogen abundance\footnote{At
  first glance, this may seem perplexing because the CCD\,S4 and B2
  spectra have larger total count numbers than the B1 spectrum from
  CCD\,S3 which could be used instead to derive the nitrogen
  abundance.  The cause of this apparent conflict originates in the
  reduced sensitivity of the front-illuminated\,(FI) CCD\,S4 at
  $\sim$0.5 keV, the energy of the N~{\sc vii} line, which is a few
  times smaller than that of the back-illuminated\,(BI) CCD\,S3.  This
  is clearly illustrated by the differing shapes of the spectra
  detected by the BI CCD\,S3 and the FI CCD\,S4 (Figure~4). As a
  result, the FI CCD\,S4 detects a count number at the N~{\sc vii}
  line which is insufficient for a reliable estimate of the nitrogen
  abundance.}, and therefore we fixed its value to that found for the
spectrum of the CCD\,S3.

The model of the X-ray-emitting plasma detected in the western regions of 
NGC\,6888 (the ACIS CCD\,S4 spectrum) has a lower dominant temperature of 
1.2$\times$10$^6$~K with a second component of 7.4$\times$10$^6$~K. 
The total observed flux is $(8.8\pm0.8)\times10^{-13}$~erg~cm$^{-2}$~s$^{-1}$, 
while the unabsorbed flux is $(1.5\pm0.3)\times10^{-11}$~erg~cm$^{-2}$~s$^{-1}$, 
with an unabsorbed flux ratio of $F_{1}/F_{2}\sim31$.

The resultant spectra from A2 and B2 are consistent from that obtained
for the total spectrum extracted from
CCD\,S4. $T_{1}=1.2\times10^{6}$~K and $T_{2}=7.4\times10^{6}$~K for
region A2, and $T_{1}=1.4\times10^{6}$~K and $T_{2}=7.5\times10^{6}$~K
for region B2\footnote{Due to the low count number of the B2 spectrum,
  the temperature of the hottest component in this region could not be
  fitted.  Consequently, the temperature of this component was fixed
  at 0.65~keV.}. The unabsorbed flux ratios are $F_{1}/F_{2}\sim$35
and $\sim26$ for A2 and B2, respectively.

\subsection{Global Properties of the hot gas in NGC\,6888}

To assess the global properties of the hot gas in NGC\,6888, we have 
derived the parameters of the two-temperature plasma emission model that 
best describes the total X-ray emission detected by \emph{Chandra}.  
The CCD\,S3 and CCD\,S4 spectra have been fitted simultaneously 
using the same hydrogen column density and plasma parameters 
($kT_1$, $kT_2$, $X_\mathrm{N}$).  
The results of this joint fit are given in Table~1 as CCD S34. 
The normalization factor of the cold thermal component of both spectra 
($A_\mathrm{1}^\mathrm{S3}$, $A_\mathrm{1}^\mathrm{S4}$) were obviously 
allowed to vary to account for the different volume emission measure of 
hot gas mapped by each detector.  
The ratio between the normalization factors of the cold and hot 
components was also allowed to vary, but it was kept the same for 
both spectra 
($A_\mathrm{2}^\mathrm{S3}$/$A_\mathrm{1}^\mathrm{S3}$ $\equiv$ 
$A_\mathrm{2}^\mathrm{S4}$/$A_\mathrm{1}^\mathrm{S4}$ $\equiv$ 
$A_\mathrm{2}^\mathrm{S34}/A_\mathrm{1}^\mathrm{S34}$), i.e., it 
was assumed that the relative contribution of both components of 
the X-ray-emitting plasma is the same across the nebula.  
The values of the normalization factors listed for this fit 
have been obtained by adding the normalization factor of 
each component for the CCD S3 and CCD S4 spectra 
($A_\mathrm{1}^\mathrm{S34}=A_\mathrm{1}^\mathrm{S3}+A_\mathrm{1}^\mathrm{S4}$, 
$A_\mathrm{2}^\mathrm{S34}=A_\mathrm{2}^\mathrm{S3}+A_\mathrm{2}^\mathrm{S4}$).  
The temperature of the two components of this model, 
$T_1=$1.4$\times$10$^6$~K and $T_2=$7.4$\times$10$^6$~K, 
are enclosed within those obtained for the CCD\,S3 and 
S4 spectra.
For comparison with all regions, we show in Figure~5 the temperature
distribution obtained from our fits for the different regions. 
Accounting for the uncertainties, there is a good agreement
for the temperature components of the different regions.  

\begin{figure}[t]
\begin{center}
\includegraphics[angle=0,width=1.05\linewidth]{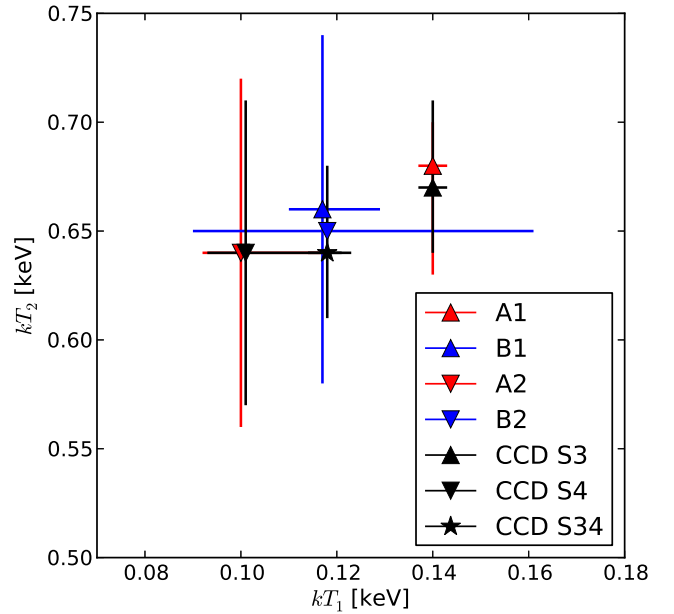}
\caption{
Plot of the temperatures of the cold and hot plasma components for the 
different spatial regions defined for the analysis of NGC\,6888.  
The inset shows a chart of the correspondence between symbols and spatial 
regions, where the star marks the location of the temperatures derived 
from the joint fit CCD\,S34.  }
\end{center}
\label{fig:contornos}
\end{figure}

The total observed flux is
$f_{\mathrm{X}}=(1.5\pm0.2)\times10^{-12}$~erg~cm$^{-2}$~s$^{-1}$,
which corresponds to an unabsorbed flux of
$F_{\mathrm{X}}=(2.5\pm0.3)\times10^{-11}$~erg~cm$^{-2}$~s$^{-1}$. The
comparison between \textit{ROSAT} PSPC and \textit{Chandra} ACIS
images indicates that the latter includes $\simeq62$~\% of the total
X-ray flux of NGC\,6888. With this we can estimate a total X-ray
intrinsic flux of
$F_{\mathrm{X,TOT}}=(4.05\pm0.5)\times10^{-11}$~erg~cm$^{-2}$~s$^{-1}$.
Adopting a distance of 1.26~kpc, the total X-ray luminosity of
NGC\,6888 in the 0.3-2.0 keV energy range is
$L_{\mathrm{X}}=(7.7\pm0.1)\times10^{33}$~erg~s$^{-1}$.

\begin{sidewaystable}
\label{tab:spectral}
\centering
\caption{Spectral Fits of the Diffuse X-ray Emission from NGC\,6888}
\scriptsize
\begin{tabular}{lcccccccccccl}
\hline
\hline
Region  & Counts      & $X_{\mathrm{N}}$ & $kT_{1}$    & $A_{1}$       & $f_{1}$                  &$F_{1}$                  & $kT_{2}$     & $A_{2}$       & $f_{2}$                &  $F_{2}$                 &  $F_{1}/F_{2}$ & $\chi^{2}$/DoF \\
        &             &                  & (keV)       &(cm$^{-5}$)    &(erg~cm$^{-2}$~s$^{-1}$)  &(erg~cm$^{-2}$~s$^{-1}$) &(keV)         & (cm$^{-5}$)   &(erg~cm$^{-2}$~s$^{-1}$)& (erg~cm$^{-2}$~s$^{-1}$) & \\      
\hline
CCD\,S3 & 14000$\pm$250 & 5.3$^{+0.7}_{-0.7}$& 0.140$^{+0.003}_{-0.003}$ & 6.1$\times10^{-3}$& 4.7$\times10^{-13}$& 7.5$\times10^{-12}$ & 0.67$^{+0.04}_{-0.03}$ & 1.7$\times10^{-4}$& 1.7$\times10^{-13}$ & 5.2$\times10^{-13}$ &  14.4 & 1.5=170.3/109 \\
A1      & 10250$\pm$200 & 5.3$^{+0.9}_{-0.9}$& 0.140$^{+0.008}_{-0.005}$ & 3.9$\times10^{-3}$& 3.2$\times10^{-13}$& 5.0$\times10^{-12}$ & 0.68$^{+0.02}_{-0.05}$ & 1.2$\times10^{-4}$& 1.2$\times10^{-13}$ & 3.7$\times10^{-13}$ &  13.6 & 1.6=173.5/105 \\
B1      & 3300 $\pm$120 & 3.7$^{+1.3}_{-1.0}$  & 0.117$^{+0.012}_{-0.007}$ & 3.2$\times10^{-3}$& 1.2$\times10^{-13}$& 2.5$\times10^{-12}$ & 0.66$^{+0.08}_{-0.08}$ & 3.9$\times10^{-5}$& 3.9$\times10^{-14}$ & 1.2$\times10^{-13}$ &  20.7 & 1.1=93.0/89 \\
\hline
CCD\,S4 & 4700$\pm$350  & 5.0                & 0.101$^{+0.020}_{-0.008}$ & 2.7$\times10^{-2}$& 7.0$\times10^{-13}$& 1.5$\times10^{-11}$ & 0.64$^{+0.07}_{-0.07}$ & 1.8$\times10^{-4}$& 1.8$\times10^{-13}$ & 4.7$\times10^{-13}$ &  30.9 & 1.5=136.6/90 \\
A2      & 3200$\pm$150  & 5.0                & 0.100$^{+0.018}_{-0.008}$ & 2.2$\times10^{-2}$& 5.5$\times10^{-13}$& 1.4$\times10^{-11}$ & 0.64$^{+0.08}_{-0.08}$ & 1.4$\times10^{-4}$& 1.3$\times10^{-13}$ & 4.2$\times10^{-13}$ &  34.5 & 1.9=115.9/59\\
B2      & 1550$\pm$100  & 5.0                & 0.118$^{+0.043}_{-0.028}$ & 3.4$\times10^{-3}$& 1.5$\times10^{-13}$& 3.1$\times10^{-12}$ & 0.65                   & 3.8$\times10^{-5}$& 3.8$\times10^{-14}$ & 1.2$\times10^{-13}$ &  25.7 & 0.6=20.13/31 \\
\hline
CCD\,S34 & \dots        & 4.0$^{+0.5}_{-0.6}$& 0.118$^{+0.005}_{-0.004}$ & 3.1$\times10^{-2}$& 1.1$\times10^{-12}$& 2.4$\times10^{-11}$ & 0.64$^{+0.04}_{-0.03}$ & 3.6$\times10^{-4}$& 3.5$\times10^{-13}$ & 1.1$\times10^{-12}$ &  22.2 & 1.5=308.1/196\\
\hline
\hline
\end{tabular}
\end{sidewaystable}

To proceed to the calculation of the electron density and mass of the
X-ray-emitting plasma in NGC\,6888 we need to adopt a geometrical
model for the nebula in order to estimate the volume occupied by the
hot plasma.  It is tempting to assume an ellipsoidal geometry, as
suggested by the H$\alpha$ images, however the \emph{Chandra} ACIS-S
and \emph{ROSAT} PSPC observations of NGC\,6888 have disclosed
emission external to the H$\alpha$ shell, just inside the [O~{\sc
  iii}] skin (see Figure~2), that implies a different physical
structure. \citet{FernandezMartin2012} describe a simple morphology
for NGC\,6888 where an ellipsoidal cavity has been carved inside the
almost spherical outer optical shell.  We can estimate lower and upper
limits of the electron density of the X-ray-emitting gas by adopting
spherical and ellipsoidal geometries, respectively.  For the spherical
model, we have adopted a radius of 9\arcmin\ to obtain an rms electron
density of $n_{\mathrm{e}}=0.4 (\epsilon / 0.1)^{-1/2}$~cm$^{-3}$,
implying a mass of the X-ray-emitting gas of $m_{\mathrm{X}} = 1.7
(\epsilon / 0.1)^{1/2}$~$M_{\odot}$, where $\epsilon$ is the gas
filling factor.  For the ellipsoidal case, we have assumed semiaxes of
9\arcmin, 6\arcmin, and 6\arcmin\ to obtain an rms electron density of
$n_{\mathrm{e}}=0.6 (\epsilon / 0.1)^{-1/2}$~cm$^{-3}$ and a mass of
the X-ray-emitting gas of $m_{\mathrm{X}} = 1.2 (\epsilon /
0.1)^{1/2}$~$M_{\odot}$.

\section{DISCUSSION}

\subsection{Comparison with previous X-ray studies}

All previous X-ray analyses of the diffuse X-ray emission from NGC\,6888 
agree on the presence of a main plasma component with a temperature 
$\simeq$1.4$\times$10$^6$ K consistent with that reported in our analysis 
of the \emph{Chandra} ACIS data.  
As for the hot component, there are notable discrepancies: 
\citet{Wrigge1994} did not find evidence of this hot component using 
\emph{ROSAT} PSPC data, 
\citet{Wrigge2005} found a hot component with a temperature marginally 
higher by $\simeq$13\% than ours using \emph{ASCA} SIS observations, 
and 
\citet{Zhekov2011} reported a second temperature component significantly 
hotter than ours, $kT\geqslant2$~keV, using \emph{Suzaku} XIS data.  
The lack of hot component in the \emph{ROSAT} PSPC data can be attributed 
to the low sensitivity of this instrument to energies above 1.0 keV.  
On the other hand, the high temperature for the secondary component reported 
by \citet{Zhekov2011} originated on the high level of X-ray emission in the 
range 1.5-4.0 keV found in the \emph{Suzaku} XIS data.   
The lack of such a hard component in the spectra derived from the 
\emph{Chandra} ACIS observations is in sharp contrast with the 
\emph{Suzaku} XIS data.  
We present in Figure~6 
the combined spectrum of point 
sources projected onto the nebula in the ACIS-S3 detector.  
The spectrum of the point sources is clearly harder than that of the 
nebula and shows significant emission above 0.8 keV up to 5-7 keV.  
Its count rate is 0.0188$\sim$0.0005 cnts~s$^{-1}$, i.e., $\sim$12\% 
that of the nebular ACIS-S3 region.  
This emission can be formally fitted using an absorbed model\footnote{
For consistency with the spectral fit of the nebular emission from 
NGC\,6888, we have adopted a hydrogen column density of 3.1$\times$10$^{21}$ 
cm$^{-2}$. 
} 
comprising an optically-thin plasma emission component at 
a temperature of (7.7$\pm$1.0)$\times$10$^6$ K and a power-law 
component with photon index $\Gamma$=1.16$\pm$0.08 for a reduced 
$\chi^2$ of 62.30/63=0.99.  
The flux of this emission in the 0.3-2.0 keV is 
(4.1$\pm$0.3)$\times$10$^{-14}$ erg~cm$^{-2}$~s$^{-1}$, 
i.e., it acounts for 6\% of the nebular flux 
measured in the ACIS-S3 spectrum in this band.  
%
%
The total flux in the 0.3-9.0 keV of this component is 
(2.46$\pm$0.24)$\times$10$^{-13}$ erg~cm$^{-2}$~s$^{-1}$.  
These results demonstrate that 
the hard, 1.5-4.0 keV component detected by \emph{Suzaku} is an 
observational artifact caused by its limited spatial resolution 
($\sim$2\arcmin), which makes difficult to identify point sources 
in the field of view of NGC\,6888 and to excise their contribution 
from the diffuse emission.

\begin{figure}[t]
\begin{center}
\includegraphics[angle=0,width=1.05\linewidth]{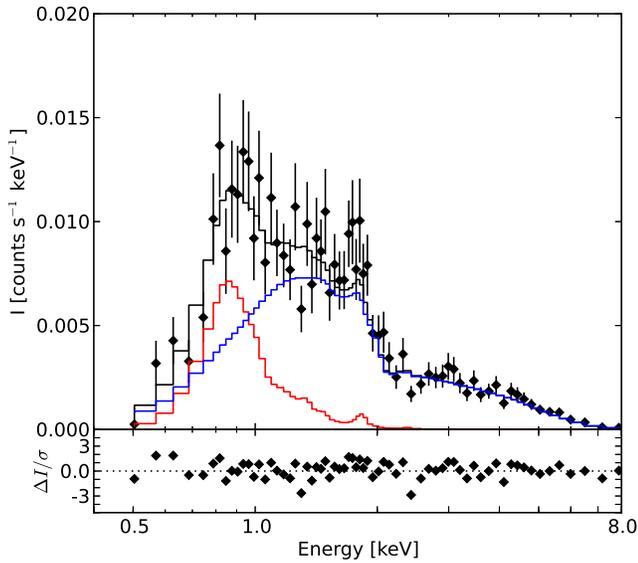}
\caption{(top panel) Background-subtracted spectrum of point sources
  projected onto the nebula NGC\,6888 in the \textit{Chandra} ACIS-S3
  detector overplotted with the best-fit model (black histogram)
  consisting of an absorbed \textit{apec} (red histogram) and a power
  law (blue histogram) model. See text for details. (Bootom panel)
  Residuals of the spectral fit.}
\end{center}
\label{fig:spec_ps}
\end{figure}

The first observations of the X-ray emission toward NGC\,6888 reported
absorbed fluxes $\sim$10$^{-12}$~erg~cm$^{-2}$~s$^{-1}$
\citep[e.g.,][]{B1988,Wrigge1994,Wrigge2005} that were raised up to
$\sim$2$\times10^{-12}$~erg~cm$^{-2}$~s$^{-1}$ by the more recent
\emph{Suzaku}'s observations \citep{Zhekov2011}. Our estimate for the
observed flux ($2.4\times10^{-12}$~erg~cm$^{-2}$~s$^{-1}$) is in good
agreement with the latest measurements despite that a fraction of the
nebula was not registered by the ACIS-S detectors and we had to rely
on \emph{ROSAT} PSPC observations to estimate the total flux.
Sensitive observations with a large field of view, as those that would
be provided by \emph{XMM-Newton}, are most needed to search for the
fainter and more extended X-ray emission in this WR nebula.

Our \emph{Chandra} ACIS-S observation yielded the detection of three
peaks in the spatial distribution of the diffuse X-ray emission in
NGC\,6888: two associated with the caps, and another one toward the NW
blowout at the western edge of the ACIS-S field of view in CCD\,S4.
The latter peak in the X-ray emission is hinted in \emph{ROSAT} PSPC
and HRI, and \emph{ASCA} SIS observations
\citep{Wrigge1994,Wrigge2002,Wrigge2005}, but it has not been reported
previously to be part of the X-ray-emitting gas associated with the
nebula\footnote{The \emph{Suzaku} observations did not map this
  nebular region, which was not registered by its detectors.  }.  This
additional emission component is located in a region with no H$\alpha$
counterpart \citep[see Figure~2;][]{Gruendl2000}, but it is spatially
delineated by the [O~{\sc iii}] outer shell.  The situation is
reminiscent of the northwest blowout in S\,308, which can be ascribed
to the action of the hot gas carving a cavity towards a low density
region of the circumstellar medium \citep[CSM;][]{Chu2003,Toala2012}.

In contrast to the limb-brightened morphology of the X-ray-emitting gas 
reported for S\,308 \citep{Toala2012}, the current (and previous) X-ray 
observations of NGC\,6888 do not show such a simple morphology, but 
that the hot gas is distributed in (at least) three maxima. 
The comparison between infrared images of NGC\,6888 and that obtained by 
\emph{ROSAT} PSPC of the X-ray-emitting gas (see Figure~6) is suggestive 
of a correlation between the regions where the X-ray emission is faintest, 
toward the southeast region of NGC\,6888, and a molecular filament traced 
in infrared wavelengths.  
Spatial variations of the amount of intervening material across the 
nebula, not accounted so far, can be playing an interesting role in 
the X-ray morphology of NGC\,6888.

Finally, it is worth mentioning here that the apparent nitrogen 
overabundances are still consistent with the values reported by
\citet{FernandezMartin2012} in their region X1.  
This region is spatially coincident with one of the brightest clumps in 
the X-ray-emitting region.

\begin{figure}[t]
\begin{center}
\includegraphics[angle=0,width=1.0\linewidth]{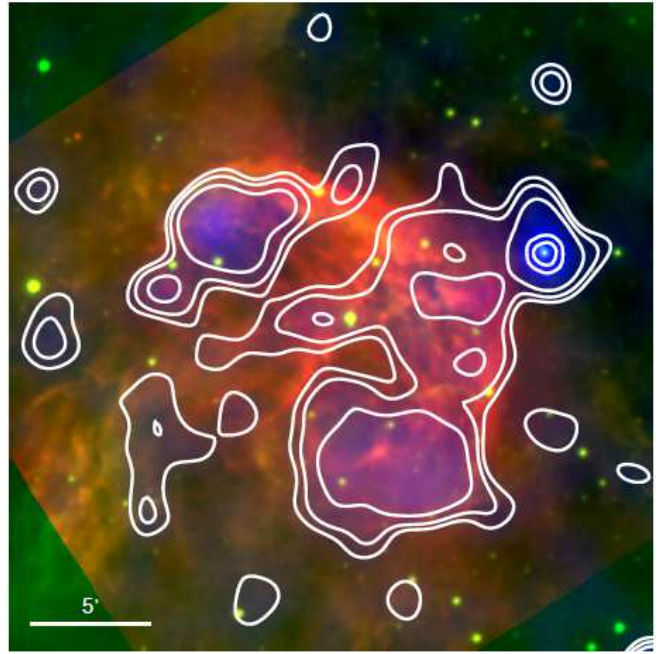}
\caption{
\emph{Spitzer} MIPS 24~$\mu$m (red), \emph{WISE} 12~$\mu$m (green), 
and \emph{ROSAT} PSPC (blue and contours) colour-composite picture 
of NGC\,6888. 
}
\end{center}
\label{fig:contornos}
\end{figure}

\subsection{Comparison with simulations}

There have been several attempts in the past to model the morphology
and X-ray emission from NGC\,6888 revealed by \emph{ROSAT},
\emph{ASCA}, and \emph{Suzaku} observations.  One-dimensional (1D)
analytic or hydrodynamic models are an elegant first approximation to
the evolution of the CSM of WR stars \citep[e.g.,][]{Zhekov1998},
although they do not include the effects of ionization due to the
central star, neither can they reproduce the wealth of structures
produced by hydrodynamical instabilities, which seem to trace the
X-ray-emitting gas in some regions inside the bubble.  Furthermore,
those simulations are not able to reproduce simultaneously regions in
which instabilities are important (e.g., the caps) and regions where
there are no clumps (the blowout). \citet{GS1996} presented 2D
hydrodynamical simulations of the evolving medium around a star with
an initial mass of 35~$M_{\odot}$. The broad morphological properties
of NGC\,6888 could be reproduced adopting a slow RSG wind of
15~km~s$^{-1}$.  These simulations were refined by \citet{Freyer2006}
who included the effects of photoionization.  Despite the progress
achieved by the 2D radiative hydrodynamic models published to date,
which have helped us advance our understanding of the formation of
NGC\,6888, they fail to produce blowout-like features.  Blowouts might
result from anisotropies in the RSG or non uniformities in the
interstellar medium.  Simulations recently presented by
\citet[][]{Rogers2013} explore the evolution of a non uniform initial
interstellar medium around massive stars.  Specific modeling
accounting for these features are needed to understand the morphology
and distribution of the X-ray-emitting gas in NGC\,6888.

It is often argued that thermal conduction between the cold
($10^{4}$~K) outer RSG material and the inner hot ($10^{7}-10^{8}$~K)
bubble causes the temperature of the hot bubble to drop to the
observed values of $T_{\noindent{X}}\sim$10$^{6}$~K
\citep[e.g.,][]{Zhekov1998,Arthur2007,Pittard2007}.  \citet{Toala2011}
recently presented numerical models including conductive effects
(classical and saturated thermal conduction) of WR bubbles that result
in a wide range of temperatures capable of generating the soft X-ray
emission observed in NGC\,6888.
Indeed, they present a spectrum corresponding to a nitrogen-rich
plasma that shows two main components, one at 0.5~keV and another at
$\sim$0.9~keV, very similar to the spectra presented in Figure~4
\citep[see also][for a comparison with models without thermal
conduction]{Dwarkadas2013}.

Finally, it is interesting to check whether the central star of
NGC\,6888 can provide the observed X-ray-emitting material.
\citet{GS1996} could match the morphology of their 2D simulations with
that of NGC\,6888 at a time $\sim$12,000~yr after the onset of the WR
phase.  That time-lapse is further reduced to $\sim$8,000~yr when the
dynamical effects on the nebular material of the photoionization are
accounted \citep{Freyer2006}.  For a mass-loss rate of
3.1$\times$10$^{-5}$ $M_\odot$~yr$^{-1}$ \citep{Abbott1986}, the total
mass provided by the star amounts up to 0.25--0.37 $M_\odot$, which is
smaller than the estimate of $>$1~$M_{\odot}$ for the mass of hot
plasma inside NGC\,6888.  Apparently, the WR wind has not had
sufficient time to inject all the hot gas inside NGC\,6888.  This
supports the possibility that physical processes such as thermal
conduction have transferred material from the outer RSG shell into the
hot bubble.  The enhanced N/O ratio of the hot plasma, similar to that
measured in the cold shell through optical emission lines, is in
concordance with this possibility.

\section{SUMMARY AND CONCLUSIONS}
We present \emph{Chandra} ACIS-S observations of the NE quadrant and
western regions of NGC\,6888. We have used these observations to study
the spatial distribution of the diffuse X-ray emission inside the
nebula and derived global values of its physical conditions. In
particular we find:

\begin{itemize}

\item The hot gas in NGC\,6888 is distributed inside the optical shell
  delineated by [O\,{\sc iii}] emission.  The spatial distribution of
  the X-ray emission shows enhancements towards the caps and a blowout
  present in the NW region of NGC\,6888. This blowout, not discussed
  in previous studies, has no H$\alpha$ counterpart, but an outer
  \emph{skin} of [O~{\sc iii}] is detected. The X-ray-emitting gas is,
  thus, traced by H$\alpha$ clumps inside the nebular shell and by the
  blowout. No clear evidence of limb-brightening is detected.

\item The X-ray emission is dominated by the N~{\sc vii} 0.5~keV line
  with additional contributions of the Fe complex and Ne~{\sc ix} line
  at 0.7--0.9~keV. The spectrum declines with energy, fading at
  energies higher than 1.5 keV. The X-ray emission from NGC\,6888 can
  be described by a two-temperature optically thin plasma emission
  model with temperatures of $\sim$1.4$\times$10$^6$~K and
  7.4$\times$10$^6$~K.

\item The intrinsic total flux emitted by NGC\,6888 in the 0.3-2 keV
  energy band is estimated to be
  $\sim(4.05\pm0.5)\times10^{-11}$~erg~cm$^{-2}$~s$^{-1}$, and the
  X-ray luminosity at a distance of 1.26~kpc is
  $L_{\mathrm{X}}=(7.7\pm0.1)\times10^{33}$~erg~s$^{-1}$.

\item The estimated rms electron density $n_{\mathrm{e}}$ of the
  X-ray-emitting gas ranges between $0.4 \times (\epsilon /
  0.1)^{-1/2}$ and $0.6 \times (\epsilon / 0.1)^{-1/2}$~cm$^{-3}$
  resulting in an estimated total mass of $1.7 \times (\epsilon /
  0.1)^{1/2} \,M_{\odot}$ and $1.2 \times (\epsilon / 0.1)^{1/2}
  \,M_{\odot}$, respectively. The density, temperature, and abundance
  of the X-ray-emitting gas are consistent with the expectation of
  thermal conduction at the wind-wind interaction zone, where the RSG
  wind material is mixed in the shocked WR wind in the bubble
  interior.

\end{itemize}

Future \emph{XMM-Newton} observations are needed to acquire a
complete view of the soft X-ray emission from the hot plasma in
NGC\,6888 with better spatial coverage, sensitivity, and energy
resolution than the existing studies. As the blowout detected in
NGC\,6888 is at the edge of the cameras of the ACIS-S instrument and a
significant section of the bubble remains unobserved,
\emph{XMM-Newton} observations could finally unveil the total
distribution of the hot gas in this nebula.  The soft sensitivity and
spatial coverage of such observations would also be very valuable to
assess the varying amounts of intervening material across the nebula
suggested by optical and infrared observations.

\acknowledgements We want to thank the anonymous referee for her/his
comments that improved the presentation of the technical details in
this paper. This work is funded by grants NASA \emph{Chandra} X-ray
observatory Guest Observer Program Grant SAO G03-4023X, and
AYA~2005-01495 of the Spanish MEC (Ministerio de Educaci\'on y
Ciencia) and AYA 2011-29754-C03-02 of the Spanish MICINN (Ministerio
de Ciencia e Innovaci\'on) co-funded with FEDER funds. JAT also
acknowledges support by the CSIC JAE-Pre student grant 2011-00189.
JAT is grateful to Y.\,Jim\'{e}nez-Teja for introducing him to the use
of matplotlib routines.

\end{document}